\documentclass[a4paper]{jpconf}

\usepackage{ifpdf}

\usepackage{graphicx}

\usepackage{latexsym}
\usepackage{amsfonts}
\usepackage{amssymb}
\usepackage{amsbsy}
\usepackage{amsmath}   %

\usepackage[usenames,dvipsnames]{pstricks}
\ifpdf
 \else
  \usepackage{epsfig}
  \usepackage{pst-solides3d}
\fi

\def\p{\partial}
\def\dfrac#1#2{{\displaystyle\frac{#1}{#2}}}
\def\stTD#1#2{\hbox to 0em{\mathsurround=0em $\stackrel{#1}{\makebox[0pt]{} #2}$\hss} \phantom{#2}}\def\stscript#1#2{\hbox to 0em{\mathsurround=0em ${\scriptstyle\stackrel{#1}{\makebox[0pt]{} #2}}$\hss} \phantom{#2}}\def\stscriptscript#1#2{\hbox to 0em{\mathsurround=0em ${\scriptscriptstyle\stackrel{#1}{\makebox[0pt]{} #2}}$\hss} \phantom{#2}}
\def\comb#1#2#3{{\mathsurround 0pt\hbox to 0pt {\hspace*{#3}\raisebox{#2}{${#1}$}\hss}}}
\def\combs#1#2#3{{\mathsurround 0pt\hbox to 0pt {\hspace*{#3}\raisebox{#2}{${\scriptstyle #1}$}\hss}}}
\def\combss#1#2#3{{\mathsurround 0pt\hbox to 0pt {\hspace*{#3}\raisebox{#2}{${\scriptscriptstyle #1}$}\hss}}}
\def\e#1{\mathrm{e}^{#1}}
\def\ii{\mathrm{i}}

\def\df{\mathrm{d}}

\def\p{\partial}
\def\dfrac#1#2{{\displaystyle\frac{#1}{#2}}}
\def\stTD#1#2{\hbox to 0em{\mathsurround=0em $\stackrel{#1}{\makebox[0pt]{} #2}$\hss} \phantom{#2}}\def\stscript#1#2{\hbox to 0em{\mathsurround=0em ${\scriptstyle\stackrel{#1}{\makebox[0pt]{} #2}}$\hss} \phantom{#2}}\def\stscriptscript#1#2{\hbox to 0em{\mathsurround=0em ${\scriptscriptstyle\stackrel{#1}{\makebox[0pt]{} #2}}$\hss} \phantom{#2}}
\def\comb#1#2#3{{\mathsurround 0pt\hbox to 0pt {\hspace*{#3}\raisebox{#2}{${#1}$}\hss}}}
\def\combs#1#2#3{{\mathsurround 0pt\hbox to 0pt {\hspace*{#3}\raisebox{#2}{${\scriptstyle #1}$}\hss}}}
\def\combss#1#2#3{{\mathsurround 0pt\hbox to 0pt {\hspace*{#3}\raisebox{#2}{${\scriptscriptstyle #1}$}\hss}}}
\def\e#1{\mathrm{e}^{#1}}
\def\ii{\mathrm{i}}

\def\fZ#1{$#1$}
\def\mphantom#1{#1}

\def\xp#1{\comb{\cdot}{-0.9ex}{0.3ex}{#1}}
\def\metr{\mathfrak{m}}
\def\metrp{\mathchoice{\comb{-}{-0.9ex}{0ex}\mathfrak{m}}{\comb{-}{-0.9ex}{0ex}\mathfrak{m}}{\combs{-}{-0.75ex}{-0.1ex}\mathfrak{m}}{}{}}
\def\nosum{\mathchoice{\textstyle \comb{\pmb{\bigl/}}{0ex}{1ex} {\sum}}{\comb{\pmb{\bigl/}}{0ex}{0.8ex} {\sum}}{}{}}
\def\eqdef{\doteqdot}
\def\EMT{\mathchoice{\combs{\to}{0.3ex}{-0.2ex}{T}}{\combs{\to}{0.3ex}{-0.2ex}{T}}{\combss{\to}{0.2ex}{-0.2ex}{T}}{\combss{\to}{0.2ex}{-0.2ex}{T}}}
\def\EMTc{\mathchoice{\combs{-}{1.5ex}{0.2ex}{\EMT}}{\combs{-}{1.5ex}{0.2ex}{\EMT}}{\combss{-}{1.1ex}{0.05ex}{\EMT}}{\combss{-}{1.1ex}{0.05ex}{\EMT}}}
\def\EMTi{\mathchoice{\combss{\infty}{1.8ex}{0.15ex}\EMT}{\combss{\infty}{1.85ex}{0.15ex}\EMT}{\combss{\infty}{1.25ex}{-0.12ex}\EMT}{\combss{\infty}{1.2ex}{-0.12ex}\EMT}}
\def\xc{\mathchoice{\comb{\boldsymbol{\cdot}}{-0.1ex}{-0.05ex}x}{\comb{\boldsymbol{\cdot}}{-0.1ex}{-0.05ex}x}{\combs{\boldsymbol{\cdot}}{-0.05ex}{-0.05ex}x}{}{}}
\def\xv{\mathchoice{\comb{\boldsymbol{\cdot}}{-0.1ex}{0.3ex}v}{\comb{\boldsymbol{\cdot}}{-0.1ex}{0.3ex}v}{\combs{\boldsymbol{\cdot}}{-0.05ex}{0.2ex}v}{}{}}
\def\pu{\mathchoice{\comb{\boldsymbol{\cdot}}{-0.1ex}{0.3ex}u}{\comb{\boldsymbol{\cdot}}{-0.1ex}{0.3ex}u}{\combs{\boldsymbol{\cdot}}{-0.05ex}{0.2ex}u}{}{}}
\def\Act{\mathcal{A}}
\def\Vol{\overline{V}}
\def\dVol{\df\mspace{-2mu}\Vol}
\def\xxx{\chi}
\def\metrEff{\mathchoice{\combs{\sim}{1ex}{0.2ex}\mathfrak{m}}{\combs{\sim}{1ex}{0.2ex}\mathfrak{m}}{\combss{\sim}{0.66ex}{0.05ex}\mathfrak{m}}{}{}}

\def\metrEffm1{\check{\metrEff}}
\def\KristffEff{\mathchoice{\combs{\sim}{1.5ex}{0.1ex}\Gamma}{\combs{\sim}{1.5ex}{0.1ex}\Gamma}{\combss{\sim}{1.15ex}{0ex}\Gamma}{}{}}
\def\Nvarphi{\mathchoice{\combs{\sim}{0.1ex}{0.05ex}{\varphi}}{\combs{\sim}{0.1ex}{0.05ex}{\varphi}}{\combss{\sim}{0.1ex}{0ex}{\varphi}}{}{}}
\def\ffun{\Phi}
\def\dffun{\ffun}
\def\LF{\mathchoice{\combs{-}{0.3ex}{-0.1ex}\mathcal{L}}{\combs{-}{0.3ex}{-0.1ex}\mathcal{L}}{\combss{-}{0.25ex}{-0.12ex}\mathcal{L}}{}{}}
\def\circffun{\ffun}
\def\circa{a}
\def\ps{\mathfrak{s}}
\def\cEp{\bar{\mathcal{E}}}
\def\cPp{\bar{\mathcal{P}}}

\begin{document}

\title[Gravitation in Theory of Space-Time Film and Galactic Soliton]{Gravitation in Theory of Space-Time Film and\\ Galactic Soliton}
\author{Alexander A. Chernitskii}

\address{$^1$ Department  of Mathematics, St. Petersburg State Chemical Pharmaceutical University, Prof. Popov str. 14, 197022 St. Petersburg, Russia}
\address{$^2$ A. Friedmann Laboratory for Theoretical Physics, The Herzen University,\\ 191186 St. Petersburg, Russia}

\ead{AAChernitskii@mail.ru}

\begin{abstract}
The scalar field of extremal space-time film is considered as unified fundamental field.
Metrical interaction between solitons-particles as gravitational interaction is considered here in approximation of a weak fundamental field.
It is shown that the signature of metrics $\{-,+,+,+\}$ in the model formulation provides the observable gravitational attraction to a region with bigger energy density of the fundamental field.
The induced gravitational interaction in the space-time film theory is applied to stars in a galaxy. The conception of galaxy soliton of space-time film is introduced.
A weak field asymptotic solution for a galaxy soliton is proposed. It is shown that the effective metrics for this solution can provide the observable velocity curves for galaxies and explains their spiral structure.
Thus a solution for so-called dark matter problem in the framework of space-time film theory is proposed.
\end{abstract}

\section{Introduction}
\label{introd}
We consider the scalar field of space-time film as unified fundamental field \cite{Chernitskii2018a}.
This scalar field model is related to well known Born -- Infeld nonlinear electrodynamics \cite{Chernitskii1999,Chernitskii2004a}.
The model under consideration is attractive because it has relatively simple and geometrically clear form.
It can be considered as a relativistic generalization of the minimal surface or
minimal thin film model in three-dimensional space.

The model under consideration provides the necessary effects which are required for a realistic unified filed model.

In particular, there are exact solutions of the model equation which can be considered as photons \cite{Chernitskii2020b}.

Also, in the space-time film theory, distant interactions of two kinds between particles-solitons,  namely force and metrical interactions \cite{Chernitskii2020c}, correspond to electromagnetic \cite{Chernitskii2017a} and gravitational \cite{Chernitskii2016b} interactions of particles respectively.

Here we apply the induced gravitational interaction in the framework of the space-time film theory \cite{Chernitskii2021a} to stars in a galaxy. So we try to solve the known so called dark matter problem.

\section{Extremal space-time film}
\label{exatf}
\begin{subequations}\label{402655631}
The following variational principle and the action which has the world volume form are considered:
 \begin{equation}
\label{35135655}
\delta \Act = 0\;,
\qquad
\Act  =\int%
_{\Vol}\!\sqrt{|\mathfrak{M}|}\;(\mathrm{d}x)^{4}
= \int%
_{\Vol}\LF\;\dVol
\;,
\end{equation}
where $\mathfrak{M} \eqdef \det(\mathfrak{M}_{\mu\nu})$,
$\left(\df x\right)^{4} \eqdef \df x^{0}\df x^{1}\df x^{2}\df x^{3}$,
 $\Vol$ is space-time volume,  $\dVol \eqdef \sqrt{|\metr|}\;\left(\df x\right)^{4}$ is four-dimensional volume element,
  \begin{equation}
 \label{381936901}
 \mathfrak{M}_{\mu\nu} = \metr_{\mu\nu} + \xxx^2\,\frac{\p \ffun}{\p x^{\mu}}\,\frac{\p \ffun}{\p x^{\nu}}
 \;,
 \qquad
 \LF  \eqdef
\sqrt{\left|1 + \xxx^{2}\,\metr^{\mu\nu}\,\frac{\p \ffun}{\p x^{\mu}}\,\frac{\p \ffun}{\p x^{\nu}}\right| }
 \end{equation}
 $\metr_{\mu\nu}$ are components of metric tensor for flat four-dimensional space-time,
 $\ffun$ is scalar real field function,
 $\xxx$ is dimensional constant.
 The Greek indices take values $\{0,1,2,3\}$.
The tensor $\mathfrak{M}_{\mu\nu}$ used here  can be called also as world tensor.
\end{subequations}

The model (\ref{402655631}) can be considered as a relativistic generalization
of the appropriate expression for the mathematical model of two-dimensional minimal thin film
in the tree-dimensional space of our everyday experience.

We consider two possible signatures of metrics in the expression of Lagrangian (\ref{381936901}):  \fZ{\{+,-,-,-\}} and \fZ{\{-,+,+,+\}}.
But an analysis of the induced gravitational interaction in the framework of this model admits the signature \fZ{\{-,+,+,+\}} only in (\ref{381936901}) for a weak fundamental field approximation.
This is shown below.

We have the following symmetrical canonical energy-momentum density tensor of the model in
Cartesian coordinates
 \begin{equation}
 \label{442613621}
\EMTc^{\mu\nu}   = \frac{1}{4\pi}\left(\frac{\dffun^{\mu}\,\dffun^{\nu}}{\LF}-
\frac{\metrp^{\mu\nu}}{\xxx^2}
\,
\LF
\right)
\;,\quad
\ffun^{\alpha}  \eqdef \metrp^{\alpha\beta}\,\frac{\p \ffun}{\p x^{\beta}}
 \;.
 \end{equation}
 where $\metrp^{\mu\nu}$ is Minkowski metrics.

To use finite integral characteristics of solutions in infinite space-time we introduce regularized energy-momentum density tensor with the following formula:
 \begin{equation}
 \label{809745461}
\EMT^{\mu\nu}   = \EMTc^{\mu\nu} - \EMTi^{\mu\nu}
\,,\qquad
\EMTi^{\mu\nu} = -\frac{1}{4\pi\,\xxx^2}\,\metrp^{\mu\nu}
 \;.
 \end{equation}
where $\EMTi^{\mu\nu}$ is the constant regularizing symmetrical energy-momentum density tensor.

\begin{subequations}\label{419746751}
The field equation in Cartesian coordinates can be written in the following form:
 \begin{equation}
\label{45213523}
\EMTc^{\mu\nu}
 \frac{\p^{2}\,\ffun}{\p x^{\mu}\,\p x^{\nu}}  = 0
 \;,
 \end{equation}
 where $\EMTc^{\mu\nu}$ is the canonical energy-momentum density tensor (\ref{442613621}).

By multiplying (\ref{45213523}) to $(-4\pi\,\xxx^{2}\,\LF)$, the field equation can be written in the following form without radicals:
 \begin{equation}
 \label{371394071}
 \left(\metrp^{\mu\nu}\left(1 + \xxx^{2}\,\metrp_{\sigma\rho}\,\ffun^{\sigma}\,\ffun^{\rho}\right) - \xxx^{2}\,\dffun^{\mu}\,\dffun^{\nu}\right)
 \frac{\p^{2}\,\ffun}{\p x^{\mu}\,\p x^{\nu}}  = 0
 \;.
 \end{equation}
\end{subequations}

 Equation (\ref{419746751}) transforms to ordinary linear wave equation with \fZ{\xxx = 0}:
 \begin{equation}
 \label{414686341}
\mphantom{
 \metrp^{\mu\nu}
 \frac{\p^{2}\,\ffun}{\p x^{\mu}\,\p x^{\nu}}  = 0
\;.
}
 \end{equation}

The model under consideration have the following notable characteristic equation:
\begin{equation}
\label{582613611}
\metrEff^{\mu\nu}\,\frac{\partial {\cal S}}{\partial x^\mu}\,\frac{\partial {\cal S}}{\partial x^\nu} = 0
\;,\quad\quad\quad
\metrEff^{\mu\nu}  \eqdef -4\pi\,\chi^2\,\EMTc^{\mu\nu}
\;,
\end{equation}
where $\metrEff^{\mu\nu}$ is the effective Riemann metrics which is proportional to the canonical energy-momentum tensor $\EMTc^{\mu\nu}$.
Equation ${\cal S}= 0$ gives a three-dimensional characteristic hypersurface of the field model in four-dimensional space-time.

\section{Gravitation as metrical interaction of solitons-particles}
\label{grav}
We consider the solitons-particles of two kind. Luminal solitons-particles have the speed of light and  subluminal solitons-particles have a subluminal speed.
For a subluminal soliton, there is an intrinsic coordinate system in which the soliton rests as a whole. A  subluminal soliton has a form of standing wave in its intrinsic coordinate system in general case.

There are also such space localized solutions for the linear wave equation (\ref{414686341}). We call these solutions weak solitons.
For example, we have the following elementary oscillating solution of the linear wave equation in the spherical coordinate system:
\begin{equation}
\label{413418791}
\mphantom{
\circffun = \frac{\circa}{\xp{r}}\sin(\xp{\omega}\,\xp{r})\,\sin(\xp{\omega}\,\xp{x}^{0})
\;,
}
\end{equation}
where \fZ{\circa} is an amplitude of the weak soliton, \fZ{\xp{\omega}} is its intrinsic frequency, the the point under symbols denotes the belonging to the intrinsic coordinate system.

For an arbitrary Cartesian coordinate system, the wave vector \fZ{\{k_\mu\}} of a subluminal soliton satisfies the following dispersion relation:
\begin{equation}
\label{66874556}
\mphantom{
\bigl|\metrp^{\mu\nu}\,k_{\mu}\,k_{\nu}\bigr|
=  \xp{\omega}^2
\;.
}
\end{equation}
And
the dispersion relation (\ref{66874556}) of a test subluminal soliton  is modified under the influence of distant solitons-particles:
\begin{equation}
\label{77874557}
\mphantom{
\bigl|\metrEff^{\mu\nu}\,k_{\mu}\,k_{\nu}\bigr|  =  \xp{\omega}^2
\;,
}
\end{equation}
where \fZ{\metrEff^{\mu\nu}} is effective metrics (\ref{582613611}) with the substitution of the field of distant solitons.
Thus here we have   \fZ{\metrEff^{\mu\nu} {}={} \metrEff^{\mu\nu}(x^\rho)} and \fZ{\bar{k}_\mu {}={} \bar{k}_\mu (x^\rho)}.

The modified dispersion relation (\ref{77874557}) gives the trajectory equation for the subluminal soliton in the following geodesic line equation form \cite{Chernitskii2021a}:
\begin{equation}
\label{564676361}
\mphantom{
\dfrac{\df \pu^\mu}{\df \ps} {}+{} \KristffEff^\mu_{\nu\rho}\,\pu^\nu\,\pu^\rho}
= 0
\;,
\qquad
\mphantom{\pu^\mu\eqdef \dfrac{\df \xc^\mu}{\df \ps}}
\;,
\qquad
\mphantom{
\KristffEff^\mu_{\nu\rho} \mphantom{\eqdef
\dfrac{1}{2}\,\metrEff^{\mu\delta}\left(
\dfrac{\p\metrEffm1_{\delta\nu}}{\p x^\rho} {}+{}
\dfrac{\p\metrEffm1_{\delta\rho}}{\p x^\nu} {}-{}
\dfrac{\p\metrEffm1_{\nu\rho}}{\p x^\delta}
\right)}
}
\;,
\end{equation}
where \fZ{\xc^j= \xc^j(x^0)} is a trajectory of energy center for the subluminal soliton, \fZ{\xc^0= x^0}, \fZ{\metrEffm1_{\mu\nu}} is the inverse tensor to the tensor \fZ{\metrEff^{\mu\nu}}.
Here the parameter of movement \fZ{\ps} satisfies the relation
\begin{equation}
\label{463511561}
\mphantom{
\df \ps^2 {\,}={\,}
\left|\metrEffm1_{\mu\nu}\,\df\xc^\mu\,\df\xc^\nu\right|}
\;.
\end{equation}

In small-speed approximation for test particle \fZ{\xv \ll 1} the geodesic line equation (\ref{564676361}) has the form
\begin{equation}
\label{Eq:Newton}
\frac{\df \xv^{i}}{\df x^{0}}  = - \KristffEff^i_{00}
\;,\qquad
\xv^{i}\eqdef \frac{\df \xc^i}{\df x^0}
\end{equation}
where $\xv_{i}$ are components of three-dimensional velocity for test soliton-particle. The Latin indices take values $\{1,2,3\}$.

In the approximation for a weak distant solitons field,  we can represent the effective metrics with the power series form in square of the nonlinearity constant \fZ{\xxx^2}
with keeping the linear terms only:
\begin{subequations}\label{380771011}
\begin{align}
\label{380812981}
\metrEff^{\mu\nu} &\approx \metrp^{\mu\nu} -
\xxx^2\left(\ffun^{\mu}\,\ffun^{\nu} - \frac{1}{2}\,\metrp^{\mu\nu}\,\metrp_{\alpha\beta}\,\ffun^{\alpha}\,\ffun^{\beta}\right)
\;,\\
\label{380812982}
\metrEffm1_{\mu\nu} &\approx \metrp_{\mu\nu} +
\xxx^2\left(\ffun_{\mu}\,\ffun_{\nu} - \frac{1}{2}\,\metrp_{\mu\nu}\,\metrp^{\alpha\beta}\,\ffun_{\alpha}\,\ffun_{\beta}\right)
\;.
\end{align}
\end{subequations}

\begin{subequations}\label{661498491}
In this approximation, we have the following expression for \fZ{\KristffEff^i_{00}} via the derivatives of fundamental field \fZ{\ffun}:
\begin{align}
\label{647214011}
\KristffEff^i_{00} &\approx
\metrp^{ii}\left(\frac{\p \metrEffm1_{0i}}{\p x^0} - \frac{1}{2}\,\frac{\p \metrEffm1_{00}}{\p x^i}\right)
 \approx
\metrp^{ii}\,4\pi\,\xxx^2\left(\frac{\p \cPp^{i}}{\p x^0} - \frac{1}{2}\,\frac{\p \cEp}{\p x^i}\right)
\qquad \left(\nosum\right)
\;,
\\
\label{67292426}
 &\cEp \eqdef \frac{1}{8\pi}\left(\left(\ffun_{0}\right)^2 + \left(\ffun_{1}\right)^2 + \left(\ffun_{2}\right)^2 + \left(\ffun_{3}\right)^2\right)\;,
 \quad
 \cPp^{i}\eqdef \frac{1}{4\pi}\,\ffun_{0}\ffun_{i}
 \;,
\end{align}
where \fZ{\cEp} and \fZ{\cPp} are energy and momentum densities for the linearized model (\ref{414686341}).
\end{subequations}

As it can be seen from (\ref{Eq:Newton}) and (\ref{661498491}), we have the attraction of test particle-soliton to a region with bigger energy density for the case
\fZ{\metrp^{ii} = 1}. Thus to have the observable gravitational attraction in this model we must choose the signature of metric \fZ{\{-,+,+,+\}} in assumption formula (\ref{402655631})
for the weak fundamental field approximation.

\section{About solving the dark matter problem}
\label{solsmp}
So called the dark matter problem has appeared from the discrepancy of available theoretical and observable velocity curves for galaxies.
The velocity curves for galaxies are the dependence of rotational velocity for stars of a galaxy from the distance to its center.

In the case of circular orbit and gravitational potential \fZ{\Nvarphi}, which is proportional  to inverse radius, we obtain the dependence
of the orbital velocity for a star from the distance to a massive galaxy center:
\begin{subequations}\label{337533851}
\begin{align}
\label{337713411}
 \Nvarphi  &= - G\,\frac{M}{r}
\;,\\
\label{337713412}
\frac{\xv^2}{r}  &= G\,\frac{M}{r^2}\quad \Longrightarrow \quad \xv =\sqrt{G\,\frac{M}{r}}
\;.
\end{align}
\end{subequations}

But the observable velocities of distant stars in a galaxy has a very small dependence from the distance to its center.

If we assume that each visible star in a galaxy has the potential of type (\ref{337713411}), we can calculate the summary gravitational potential of galaxy.
But in this case we also do not obtain the observable rotation curves. For a solution of the problem we can assume an existence of a dark matter or
we can abandon the galaxy potential which is obtained by ordinary summation of star potentials of type (\ref{337713411}).

The conception of induced gravitation in the framework of the space-time film theory gives a radical solution of this problem.

All observable galaxies has a disk form, usually with helicoidal structure. Thus they have a certain axis symmetry. In the point of view of
 the space-time film theory we must have an appropriate galaxy solution or galaxy soliton or soliton-galaxy. Far from the region of space with high
 energy density, we can assume that the fundamental field \fZ{\ffun} satisfies the linearized equation of the model or ordinary linear wave equation (\ref{414686341}).
 It is natural to consider in this connection its solutions with the appropriate galaxy symmetry.

As the possible reductive  asymptotics of the galactic soliton, let us consider the following solution of the linear wave equation (\ref{414686341}) in cylindrical coordinate system \fZ{\{\rho,\varphi,z\}}:
\begin{equation}
\label{758662921}
\ffun  =
\Re\Biggl[
\biggl(J_{m}\Bigl(\sqrt{\lambda^2 + \omega^2}\,\rho\Bigr)
+ \ii\,Y_{m}\Bigl(\sqrt{\lambda^2 + \omega^2}\,\rho\Bigr)\biggr)
\,\e{-\lambda \left|z\right| + \ii\bigl(m\,\varphi - \omega\,x^{0}\bigr)}
\Biggr]
\;.
\end{equation}
where \fZ{J_{m}} and \fZ{Y_{m}} are Bessel functions of the first and second kinds, \fZ{J_{m} + \ii\,Y_{m}} is the Hankel function of the first kind.

\begin{figure}[h]
\ifpdf
\includegraphics[width=20pc]{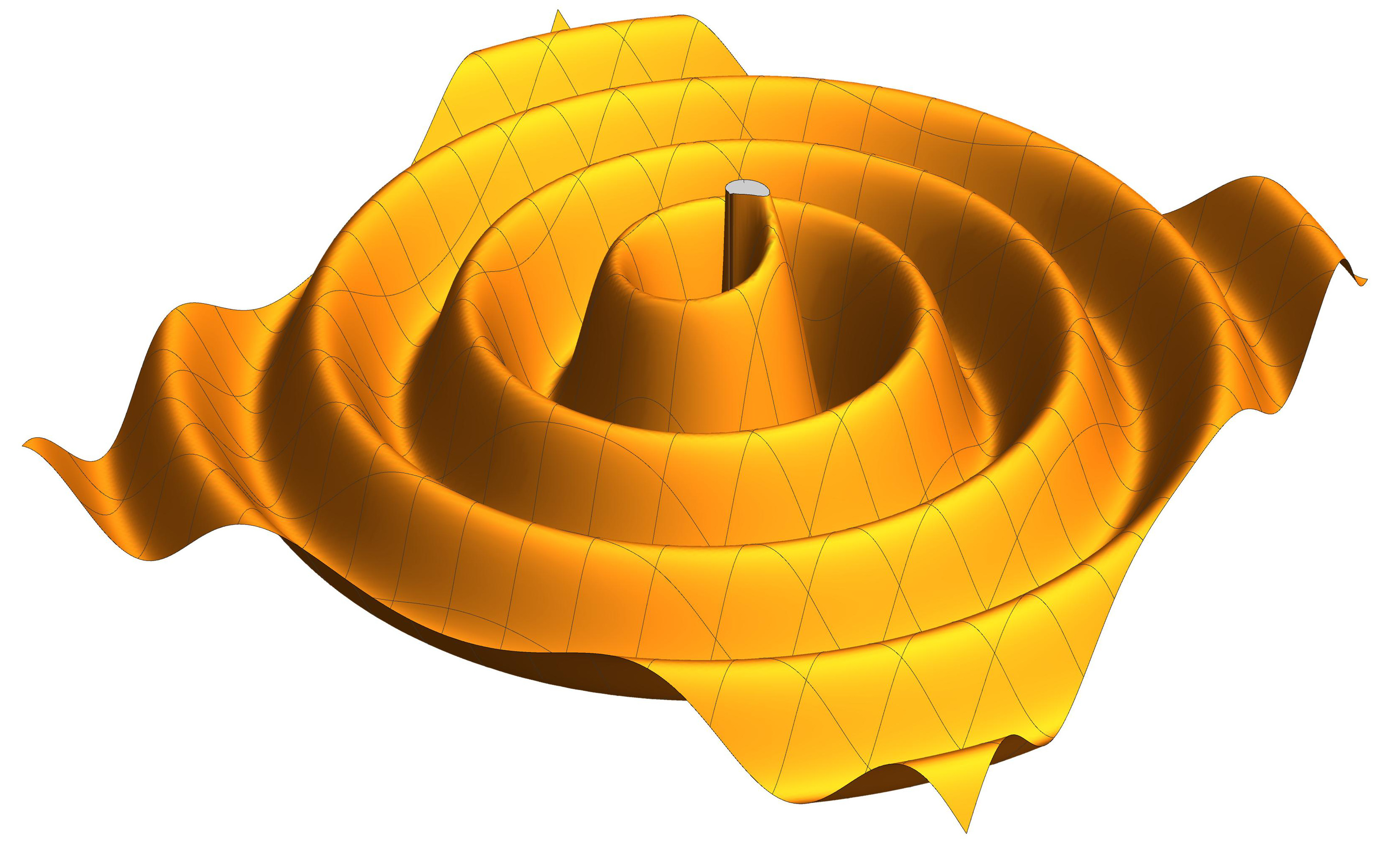}\hspace{2pc}%
\else
\includegraphics[width=20pc]{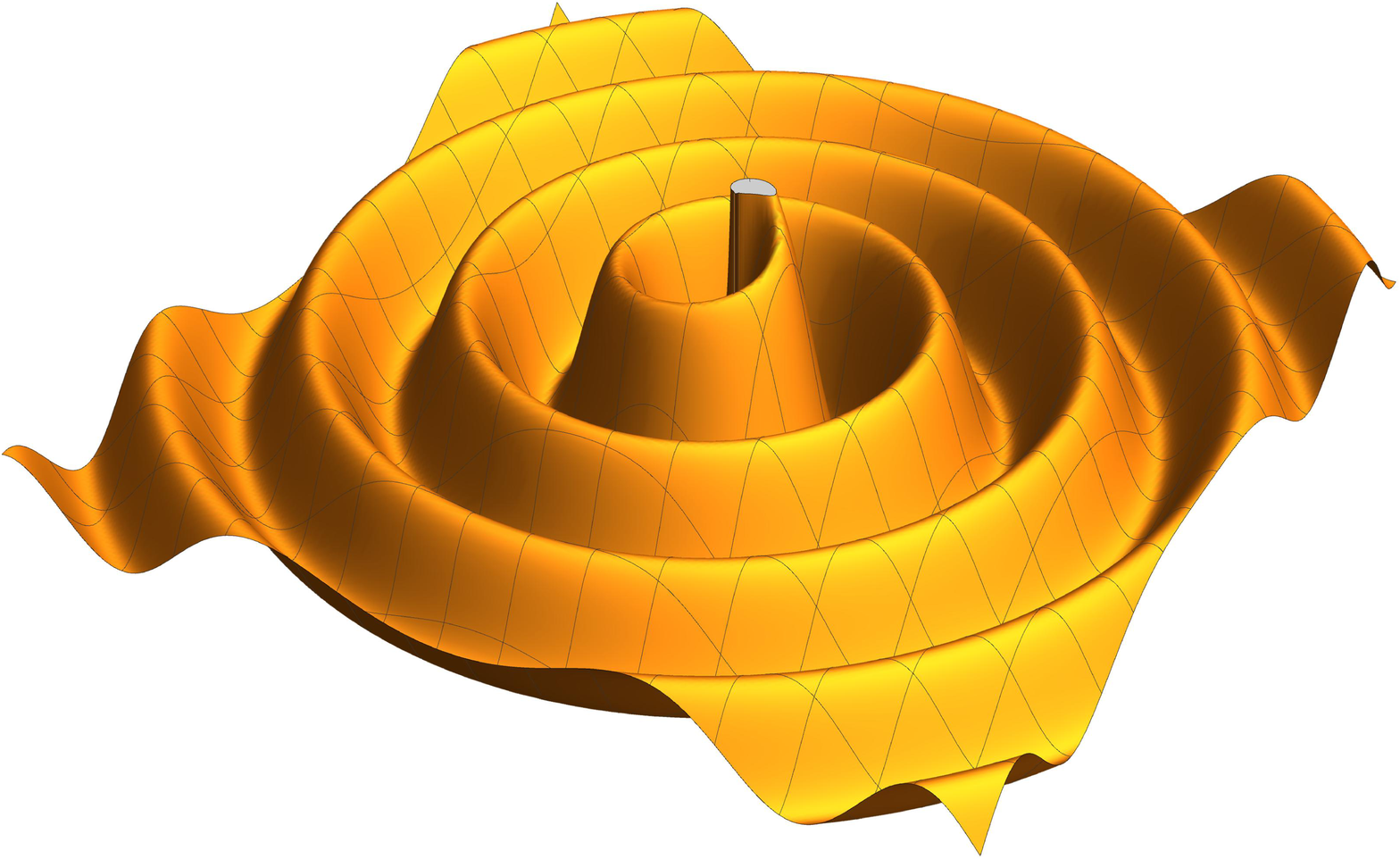}\hspace{2pc}%
\fi
\begin{minipage}[b]{15.5pc}\caption{\label{labelFigB}Section \fZ{(z=0) \cap (x^{0} = 0)}  of the possible simplest asymptotics (\ref{758662921})  (\fZ{m=1}, \fZ{\omega=\lambda=1}) for a galactic soliton.}
\end{minipage}
\end{figure}

The solution (\ref{758662921}) has the singular plane \fZ{z = 0} and the singular axis \fZ{\rho = 0}.
It is a rotating spiral configuration with exponentially decreasing amplitude of the field for \fZ{z\to\pm\infty}.
The section \fZ{(z=0) \cap (x^{0} = 0)}  of the solution (\ref{758662921}) for \fZ{m=1}, \fZ{\omega=\lambda=1} is shown on Fig. \ref{labelFigB}.

\begin{subequations}\label{817187051}
In view of the equation of motion (\ref{Eq:Newton}) and expression for \fZ{\KristffEff^i_{00}} (\ref{661498491}) we obtain the orbital velocity
for an asymptotic function \fZ{\ffun}:
\begin{align}
\label{728908881}
&\frac{\xv^2}{\rho}  = \KristffEff^1_{00}\quad \Longrightarrow \quad\xv =\sqrt{\rho\,\KristffEff^1_{00}}\;,\quad
\KristffEff^1_{00} \geqslant 0
\;,\\
\label{728908882}
& \KristffEff^1_{00} \approx
\left(\frac{\p \metrEffm1_{01}}{\p x^0} - \frac{1}{2}\,\frac{\p \metrEffm1_{00}}{\p \rho}\right)
 \approx
4\pi\,\xxx^2\left(\frac{\p \cPp^{1}}{\p x^0} - \frac{1}{2}\,\frac{\p \cEp}{\p \rho}\right)
\;.
\end{align}
\end{subequations}
Here for \fZ{\KristffEff^1_{00} < 0} we have the repulsion from the center and the orbital velocity is absent.

\begin{figure}[h]
\begin{minipage}{17.7pc}
{\unitlength 1.0mm
\begin{picture}(50,75)
\put(0,0){
\ifpdf
 \put(0,0){\includegraphics[width=18pc]{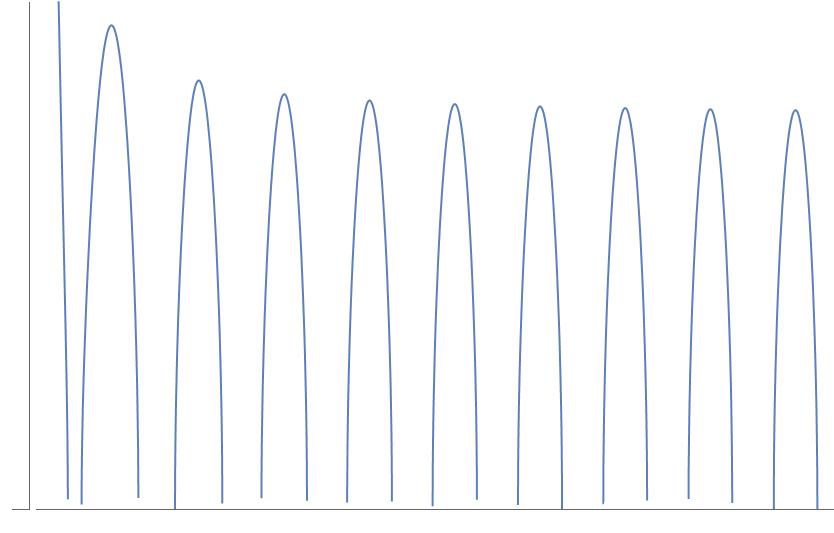}}
\else
 \put(0,0){\includegraphics[width=18pc]{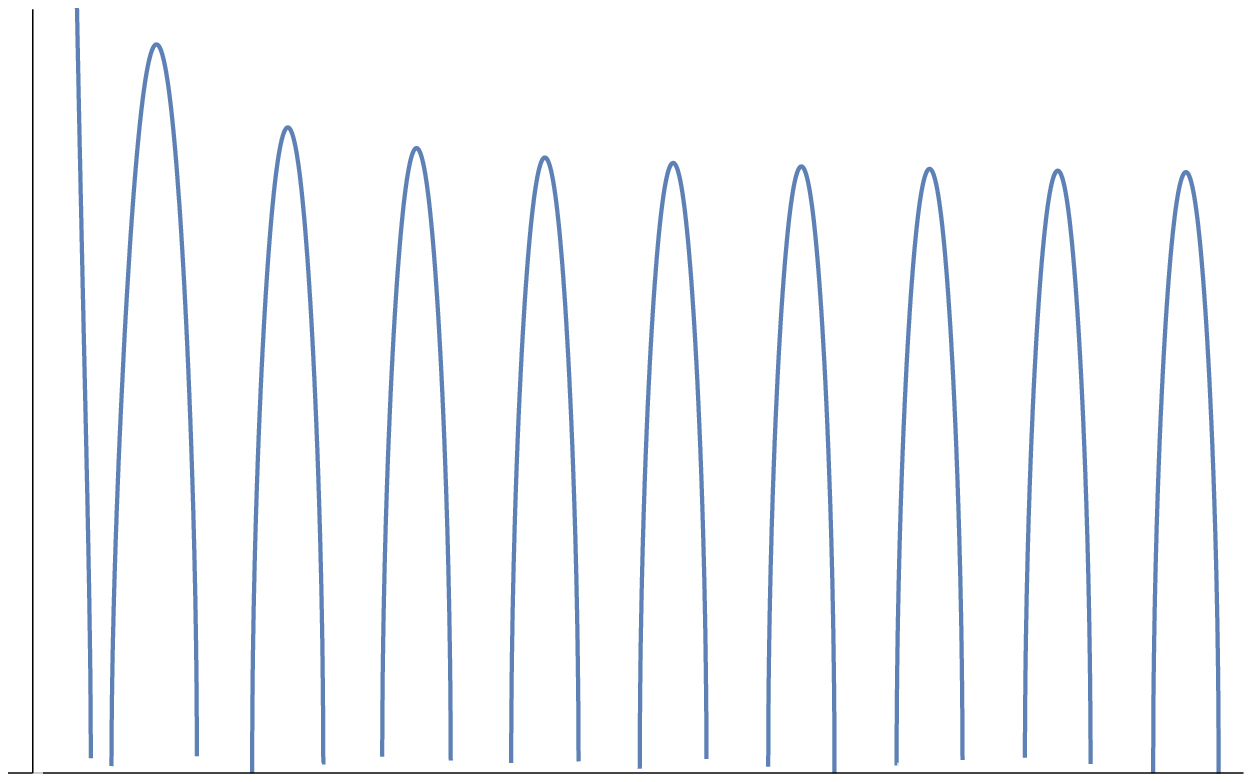}}
\fi
\put(0,50){\makebox(0,0)[lc]{$\xv$}}
\put(76,0){\makebox(0,0)[lc]{$\rho$}}
}
\end{picture}
}
\caption{\label{labelFigD} The orbital velocity for the solution (\ref{758662921}) on the line \fZ{(z=0) \cap (\varphi = 0) \cap (x^{0} = 0)} for \fZ{m=1}, \fZ{\omega=\lambda=1}.}
\end{minipage}\hspace{2pc}%
\begin{minipage}{17.7pc}
{\unitlength 1.0mm
\begin{picture}(50,75)
\put(-5,0){
\ifpdf
 \put(0,0){\includegraphics[width=20pc]{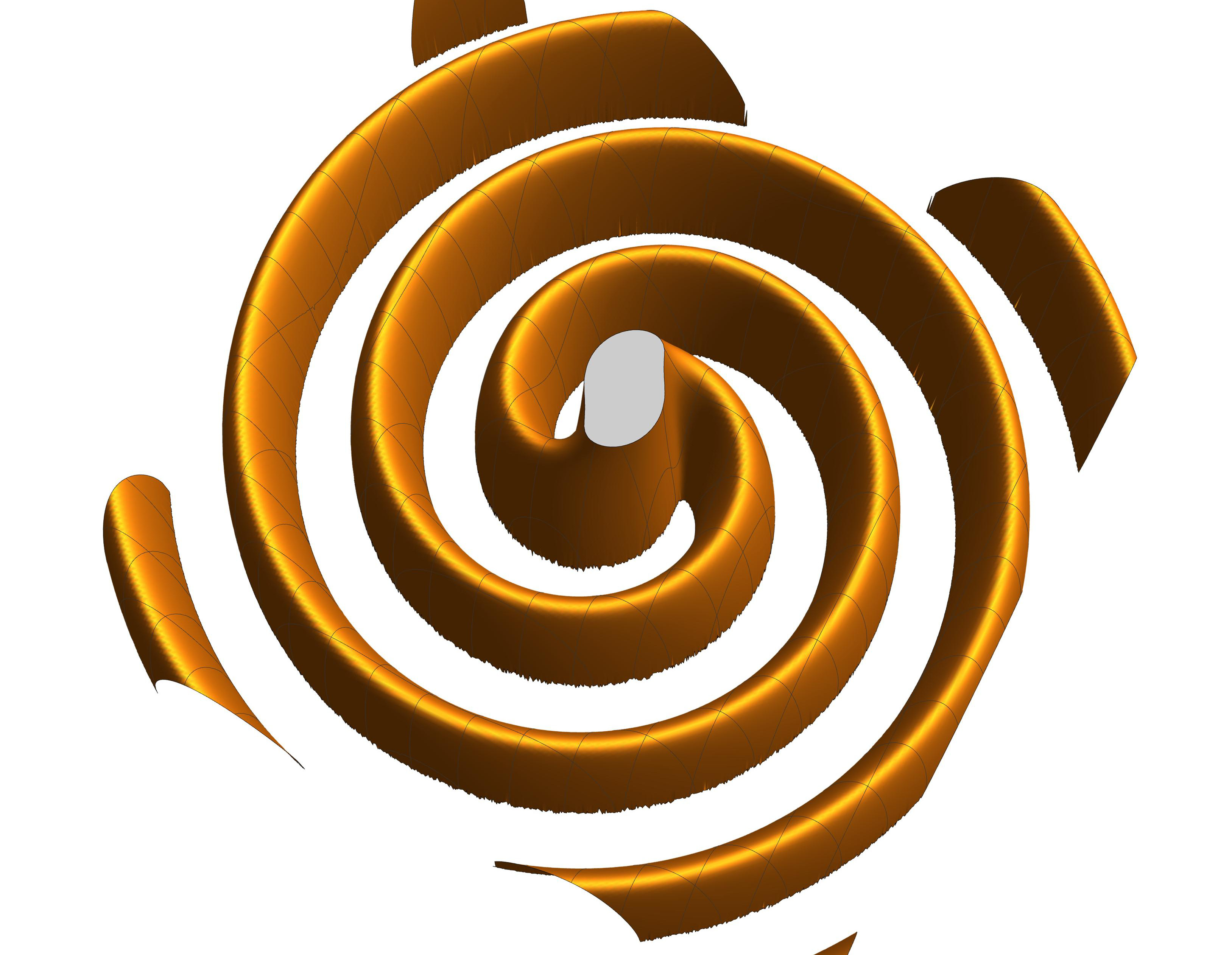}}
\else
 \put(0,0){\includegraphics[width=20pc]{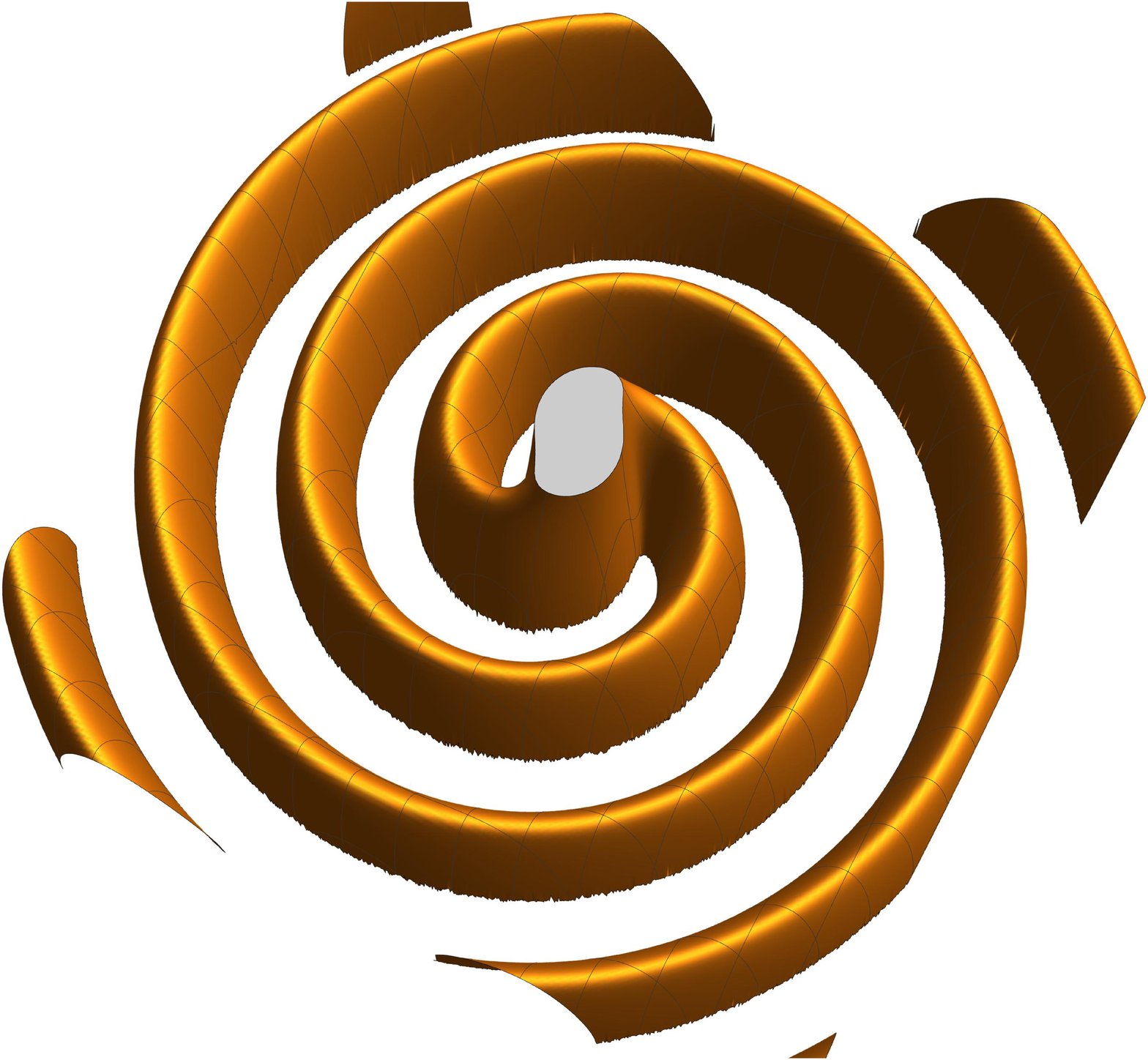}}
\fi
}
\end{picture}
}
\caption{\label{labelFigC1} The orbital velocity for the solution (\ref{758662921}) on the plane \fZ{(z=0) \cap (x^{0} = 0)} for \fZ{m=1}, \fZ{\omega=\lambda=1}.}
\end{minipage}
\end{figure}

The orbital velocity (\ref{728908881}) for the asymptotic solution (\ref{758662921}) on the line \fZ{(z=0) \cap (\varphi = 0) \cap (x^{0} = 0)} for \fZ{m=1}, \fZ{\omega=\lambda=1} is shown on Fig. \ref{labelFigD},
and on the plane \fZ{(z=0) \cap (x^{0} = 0)} for \fZ{m=1}, \fZ{\omega=\lambda=1} is shown on Fig. \ref{labelFigC1}.

As can be seen on Fig. \ref{labelFigD}, in the region of space, where the orbital velocity exists, its average value has the tendency to be constant. This
coincides with experimental data. Also, helical structure shown on Fig. \ref{labelFigC1} coincides with the images of real galaxies.

It should be noted that for the considered simple asymptotics of galaxy soliton (\ref{758662921}), its number of branches or tails equals \fZ{2\,m}, where \fZ{m} is the order of twist for the space-time film soliton. As we know, the observable galaxies have even number of branches, that confirms the considered concept of galaxy soliton of space-time film.

Here we have considered only one possible asymptotics of a galactic soliton which coincides with experimental data. But, apparently, for any real galaxy
we can discover the appropriate asymptotic solution of the space-time film equation which gives an observable distribution of star density and orbital velocities.
This will mean that for this galaxy we have obtained the asymptotics of the appropriate galactic soliton.

\section{Conclusions}
\label{concl}
Thus we have considered the effect of induced gravitation in the framework of space-time film theory.
The approximation of a weak fundamental field for the effective metrics is considered.
It is discovered that the choice for the signature of metrics \fZ{\{-,+,+,+\}} in the model formulation provides the gravitational attraction to a region with bigger energy density of the fundamental field.

We introduce the conception of galactic soliton of space-time film.
In the framework of this conception, the possible solution of so called dark matter problem and explanation of galactic spiral structure are proposed.
We can note the defined agreement of the galaxy soliton conception with the observed data.
In particular, this approach gives the observable even number of branches for a spiral galaxy.

\providecommand{\newblock}{}

\end{document}